\documentclass[aps,prl,twocolumn,floatfix,showpacs,amsmath,amssymb,nofootinbib,10pt,longbibliography]{revtex4-1}

\usepackage{mathtools}
\usepackage{bbm}
\usepackage{color}
\usepackage{graphicx}
\usepackage{IEEEtrantools}
\usepackage{txfonts}
\usepackage[pdftex]{hyperref}

\DeclareFontFamily{OT1}{pzc}{}
\DeclareFontShape{OT1}{pzc}{m}{it}{<-> s * [1.10] pzcmi7t}{}
\DeclareMathAlphabet{\mathpzc}{OT1}{pzc}{m}{it}

\begin{document}

\title{Solvable Markovian dynamics of lattice quantum spin models}

\author{D.\ Mesterh\'{a}zy}
\author{F.\ Hebenstreit}
\affiliation{Albert Einstein Center for Fundamental Physics, Institute for Theoretical Physics, University of Bern, 3012 Bern, Switzerland}

\begin{abstract}
  We address the real-time dynamics of lattice quantum spin models coupled to single or multiple Markovian dissipative reservoirs using the method of closed hierarchies of correlation functions. This approach allows us to solve a number of quantum spin models exactly in arbitrary dimensions, which is illustrated explicitly with two examples of driven-dissipative systems. We investigate their respective nonequilibrium steady states as well as the full real-time evolution on unprecedented system sizes. Characteristic timescales are derived analytically, which allows us to understand the nontrivial finite-size scaling of the dissipative gap. The corresponding scaling exponents are confirmed by solving numerically for the full real-time evolution of two-point correlation functions.
\end{abstract}

\pacs{75.10.Jm, 75.10.-b, 05.50.+q}
%Quantum spin models
%General theory and models of magnetic ordering 
%Lattice theory and statistics

\maketitle

{\it Introduction.} The dynamics of quantum spin systems is of fundamental importance to atomic, molecular, and optical physics, condensed matter physics, quantum chemistry, and beyond. Yet, computing their real-time dynamics poses a tremendous challenge for classical computers, since the required resources scale exponentially with the system size. In general, standard methods for quantum spin systems in equilibrium as, e.g., Monte Carlo importance sampling \cite{Suzuki:1977,Suzuki:1993}, cannot be applied out of equilibrium due to the absence of a positive-definite probability measure. The density matrix renormalization group \cite{White:1992}, on the other hand, which can be understood as a variational method on the class of matrix product states \cite{Verstraete:2008,Schollwock:2011}, manages to reduce the exponential complexity by following the time-dependence of quantum states with low entanglement \cite{Cazalilla:2002,White:2004,Vidal:2004,Daley:2004}. Unfortunately, this restriction largely limits its applicability to one-dimensional systems and typically does not allow to follow the time evolution of the system to arbitrarily large times. Other time-dependent variational methods \cite{Haegeman:2011,Transchel:2014} rely on similar {\it ansatz} states and hence also face similar constraints. In view of these limitations, exactly solvable models are immensely useful. Not only do they provide important insights into intriguing nonequilibrium phenomena, but they also serve as benchmarks for systematic approximations and numerical techniques.

In this Letter, we investigate quantum spin models that are coupled to a Markovian reservoir, which can be solved {\it exactly} in the framework of closed hierarchies of correlation functions \cite{Caspar:2015ddt,Caspar:2016rns} (see also earlier work Ref.\ \cite{Zunkovic:2014,Klich:2015} in the context of many-body boson and fermion systems). Their solution relies on a set of closure conditions that need to be satisfied in order for correlation functions of quantum spins (multi-point functions) to close among themselves. One is bound to ask whether these models are still sufficiently interesting to warrant further discussion -- are they maybe ``too simple to be real''? Here, we argue that the latter is {\it not} the case. Indeed, several protocols for the dissipative preparation of long-range entangled states \cite{Kraus:2008,Diehl:2008,Verstraete:2009,Weimer:2010} lie within this class of exactly solvable models, some of which have also been realized recently using ion-trap experiments \cite{Barreiro:2011,Schindler:2013}. Other engineered dissipative systems with competing dissipative couplings \cite{Weimer:2017} are also directly accessible within our approach. This gives us the unique opportunity to investigate these systems on sufficiently large system sizes in order to determine those points in parameter space where the dissipative gap closes. There, the relevant length- and timescales diverge, which is characteristic of a dissipative phase transition \cite{Verstraete:2009,Diehl:2010,Eisert:2010,Lesanovsky:2013}. In particular, we calculate the finite-size scaling of the dissipative gap, which characterizes the nature of the long-time relaxation towards the nonequilibrium steady state and essentially defines the nonequilibrium dynamical universality class. In combination with exact diagonalization of the quantum master equation on small system sizes, closed hierarchies allow us to construct a detailed picture of the full nonequilibrium dynamics. These insights might also help to elucidate the dynamics of systems that are outside of scope of these methods. In this spirit, we conclude this work with some general arguments regarding the properties of driven-dissipative quantum systems. 

{\it Closed hierarchies of correlations.} Before we turn to specific examples, we need to make the nature and properties of closed hierarchies precise. We will be concerned with open systems of quantum spin-$1/2$ particles (qubits) on a regular $d$-dimensional hypercubic lattice with periodic boundary conditions. The latter assumption is neither necessary for the following arguments, nor essential for our conclusions, but it allows for exploiting the translation symmetry of the problem. We assume that the spins are weakly coupled to an external reservoir and that there is a separation of timescales between the dynamics of the quantum spin system and the reservoir. This second assumption allows us to integrate out the ``fast'' degrees of freedom of the bath and to consider only the {\it reduced} density matrix $\rho_N$ (i.e., the statistical ensemble) of the $N$-spin system. Proceeding along these lines, we arrive at an effective description of the dynamics of the reduced system $\rho_N$, which is described by a Markovian quantum master equation \cite{Breuer:2002}.

To address the time evolution of observables $A$, which are elements of the set $\mathfrak{B}(\mathcal{H}_N)$ of bounded linear operators on the Hilbert space $\mathcal{H}_N\simeq \mathbbm{C}^{2^N}$, it is convenient to work in the Heisenberg picture, where their time evolution is governed by the (adjoint)  Markovian quantum master equation
\begin{equation}
  \frac{d}{dt} A(t) = \frac{i}{\hbar} \left[H,A\right] + \sum_{\ell} \gamma_{\ell} \boldsymbol{\mathcal{D}}_{\ell}(A) .
  \label{eqm}
\end{equation}
$H$ is the system Hamiltonian, and $\gamma_{\ell}=\gamma g_{\ell} $ defines the rate of dissipation, where $\gamma$ sets the scale and $g_{\ell}$ denotes the dimensionless dissipative coupling. In the following, we will always set $\hbar = \gamma = 1$. The ``dissipator'' $\boldsymbol{\mathcal{D}}_{\ell}$ is conveniently represented in Lindblad form \cite{Gorini:1975nb,Lindblad:1975ef}
\begin{equation}
  \boldsymbol{\mathcal{D}}_{\ell}(\hspace{0.9pt}\bullet\hspace{0.7pt}) = \ell^\dagger \hspace{0.5pt} \big[ \hspace{0.9pt}\bullet\hspace{0.7pt} , \ell\big]+\big[\ell^\dagger, \hspace{0.7pt}\bullet\hspace{0.9pt} \big] \hspace{0.9pt} \ell , 
\end{equation}
where $\ell$ are (Lindblad) jump operators that describe the effective interaction with the environment. If one considers individual realizations of the statistical ensemble $\rho_N$, these jump operators yield a stochastic time evolution of observables \cite{Gardiner:1991,Breuer:2002}.

\vskip 4pt

Specifically, we will be interested in the time evolution of multi-point functions
\begin{equation}
  A_m : \mathtt{C}_m \rightarrow \mathfrak{B}(\mathcal{H}_N) ,
  \label{map}
\end{equation}
which can be viewed as maps from the configuration space of $m$ spin-$1/2$ particles, $\mathtt{C}_m$, to the space of bounded operators on the $N$-particle Hilbert space, $\mathcal{H}_N$. We define the map \eqref{map} for $m > 0$ such that $A_m(X) = \prod_{i = 1}^m s_{x_i}^{a_i}$ for each point $X = \{ ( x_i, a_i ) \}_{1\leq i\leq m}\in \mathtt{C}_m$ (i.e., $x_i\neq x_j$, $\forall i\neq j$) and $A_0$ is identified with the identity operator. Inserting the operator $A_m(X)$ into \mbox{Eq.\ \eqref{eqm}}, we observe that the commutators on the right hand side of this equation typically induce operators of higher order, i.e.,
\begin{equation}
  i \left[H, A_m(X) \right] + \sum_{\ell} g_{\ell} \boldsymbol{\mathcal{D}}_{\ell} \left(A_m(X)\right) = c_0 + \sum_{n = 1}^{\infty} \sum_{Y\in\mathtt{C}_n} c_n(Y) A_n(Y) ,
  \label{hinf}
\end{equation}
with coefficients $c_n\in \mathbb{R}$. For most systems of interest, Eq.\ \eqref{hinf} yields an infinite hierarchy of equations of motion that cannot be solved without truncation. One may ask, however, under what conditions the hierarchy closes {\it exactly}, such that
\begin{equation}
  c_n(Y) = 0 , \quad \forall n > \mathcal{N}\!\left(A_m(X)\right) .
\end{equation}
$\mathcal{N}\!\left(A_m(X)\right)$ is an integer that, in general, might depend on the particular observable under consideration. Stated differently, we are asking the following question: Are there quantum spin systems that yield a closed hierarchy, and if yes, what are their properties? The strongest form of the closure conditions
\begin{equation}
  \mathcal{N}\!\left(A_m(X)\right)= m , \quad \forall X, 
  \label{cond}
\end{equation}
was derived by us in previous work \cite{Caspar:2016rns}. This condition constrains the form of the Hamiltonian $H$ and the jump operators $\ell$ and defines a class of efficiently solvable models (in the sense that the problem of computing the time evolution of $m$-point functions scales only polynomially in the system size $N$). The general idea behind condition \eqref{cond} can be easily summarized as follows: Inserting $m$ quantum spin operators $s_{x_i}^{a_i}$ on lattice sites $x_i$, $i = 1,2, \ldots, m$ (which defines the $m$-point operator, $A_m(X)$, of interest), the infinitesimal generator of the dynamical semigroup $\boldsymbol{\mathcal{G}}(\hspace{0.9pt}\bullet\hspace{0.7pt}) = i \left[H, \hspace{0.9pt}\bullet\hspace{0.7pt} \right] + \sum_{\ell} g_{\ell} \boldsymbol{\mathcal{D}}_{\ell} \left(\hspace{0.9pt}\bullet\hspace{0.7pt}\right)$ is allowed only {\it (i)} to move around single quantum spin operators from one site to another, {\it (ii)} to convert them into different ``species'', i.e.,  $s_x^a\mapsto s_y^b$, or {\it (iii)} to completely remove them from the lattice. Thus, systems that satisfy the closure condition \eqref{cond} are characterized by hopping of quantum spins from one site to another, with some probability for single spins being eliminated from the lattice. In a diagrammatic language, these processes can be viewed either as $m \rightarrow m-k$ particle scattering ($A_m \rightarrow A_{m-k}$, $0\leq k\leq m-1$), or decay $m\rightarrow 0$ ($A_m \rightarrow \varnothing$). The opposite process, however, namely particle creation ($A_m\rightarrow A_{m+k}$, $k > 0$) is {\it not} allowed. Therefore, if we define the set of operators $A|_{\mathtt{C}_m}\equiv A_m(\mathtt{C}_m)$, we may write \mbox{Eq.\ \eqref{cond}} in short as $\boldsymbol{\mathcal{G}}(A|_{\mathtt{C}_m}) = \bigoplus_{n = 0}^{m} c_n A|_{\mathtt{C}_n}$.

\vskip 4pt

One might imagine less restrictive conditions on the dynamics, e.g.,
\begin{equation}
  \mathcal{N}\!\left(A_m(X)\right) = m + k , \quad k \geq 0 ,
  \label{cond2}
\end{equation}
with $k$ finite. However, it is easy to see that the hierarchy is unbounded if transitions of the type $A_m \rightarrow A_{m+k}$, $k > 0$, are allowed in general. The key to construct closed hierarchies in this case is the following observation: The action of the infini\-tesimal generator $\boldsymbol{\mathcal{G}}$ should map the set of operators $A|_{\mathtt{C}_m}$ to a subset $A|_{\mathtt{S}}\subseteq \bigoplus_{n = 0}^{m + k} c_n A|_{\mathtt{C}_n}$, which itself defines an invariant subspace for the dynamics, i.e., $\boldsymbol{\mathcal{G}}(A|_{\mathtt{S}})\subseteq A|_{\mathtt{S}}$. In this case, only the knowledge of a subset of operators, namely $A|_{\mathtt{S}}$, is needed to determine the time evolution of $m$-point functions $A_m$. Finding efficiently solvable models that satisfy Eqs.\ \eqref{cond} or \eqref{cond2} hence corresponds to the problem of identifying invariant subspaces $A|_{\mathtt{S}}$ which are {\it small} (i.e., whose size should scale only polynomially with the system size). The dissipative transverse-field Ising model \cite{Foss:2017} is one example of a system that satisfies Eq.\ \eqref{cond2}, which is of interest in the context of ``engineered'' quantum systems, e.g., using ion trap experiments \cite{Garttner:2016mqj} or Rydberg atoms \cite{Weimer:2010,Glaetzle:2015,Roghani:2016}. Here, we will limit ourselves to the condition Eq.\ \eqref{cond} and discuss two examples within this class of problems that have been at the focus of interest recently.

\vskip 4pt

{\it Dissipative quantum state preparation.} We consider the set of non-Hermitian jump operators \cite{Diehl:2008}
\begin{equation}
  \ell\in  \left\{\hspace{0.5pt}L_{xy} = \frac{1}{2}\delta_{\langle x,y \rangle}(s_{x}^{+}+s_{y}^{+}) (s_x^{-}-s_y^{-}) \right\} ,
  \label{ch1}
\end{equation}
that act uniformly on adjacent sites with $g_{\ell}= 1$; $\delta_{\langle x,y\rangle}$ is nonzero and equal to one if and only if $x$ and $y$ are nearest neighbor sites, i.e., $||x-y|| = 1$ (here and in the following the lattice spacing is set to one). Note that $L_{xy}$ acts on pairs of nearest-neighbor sites $(x,y)$ and maps the two-particle spin singlet state to the spin triplet, while it conserves $S^3= \sum_x s_x^3$. At the same time it annihilates the spin triplet state, i.e., $L_{xy}^2 = 0$. 

\newpage

In the absence of a Hamiltonian, $H = 0$, we obtain a linear system of equations of motion
\begin{equation}
  \frac{d}{dt} A_m(X) = \sum_{Y\in\mathtt{C}_m} c_m(Y) A_m(Y) ,
  \label{closed}
\end{equation}
which can be solved exactly for arbitrary $m$-point functions.\footnote{Strictly speaking, the assumption $H = 0$ is not necessary since Eq.\ \eqref{cond} also allows for the possibility of including a coupling to an external field, i.e., $H = \sum_{x,a} h_x^a s_x^a$, with $h_x^a\in \mathbbm{R}$.} The dynamics of the (ensemble-averaged) two-point correlation functions $C_{xy}= \operatorname{tr} \rho_N \big( s_{x}^{+} s_{y}^{-} + s_{x}^{-} s_{y}^{+}\big)$ and $D_{xy}= \operatorname{tr} \rho_N s_{x}^{3} s_{y}^{3}$ is governed by the evolution equations ($x\neq y$)
\begin{align}
  \frac{d}{d t} C_{xy} &= \frac{1}{2}\Delta_{xy} C_{xy} -\delta_{\langle x,y\rangle} \left(C_{xy}+4 D_{xy} \right) , \label{eq1} \\
  \frac{d}{d t} D_{xy} &= \frac{1}{2}\Delta_{xy} D_{xy} -\delta_{\langle x,y\rangle} \left(D_{xx}- D_{xy} \right) .\label{eq2}
\end{align}
$\Delta_{xy}\equiv \Delta_x + \Delta_y$ is the discrete Laplacian on the hypercubic lattice and Eqs.\ \eqref{eq1} and \eqref{eq2} are subject to the boundary condition $C_{xx} = 4 D_{xx} = 1$. We choose to initialize the system in a fully disordered state with the reduced density matrix $\rho_N=2^{-N}\mathbbm{1}$, i.e.,
\begin{equation}
  C_{xy}(t = 0) = 4 D_{xy}(t = 0) = \delta_{xy} .
  \label{ini}
\end{equation}
This set up allows us to study the build-up of correlations by putting the spin system in contact with the ``engineered'' reservoir. In fact, these initial conditions are rather convenient, since Eq.\ \eqref{eq2} can be solved trivially in this case -- note that the two-point correlation function is invariant in time $D_{xy}(t) = (1/4)\delta_{xy}$ so that the growth of correlations is fully described by the function $C_{xy}$ alone.

In order to study its long-time behavior, it is useful to consider the limit of large separations $||x-y||\gg 1$, in which case the contact term on the right hand side of Eq.\ \eqref{eq1} vanishes. If we furthermore assume that the initial state $\rho_N$ is translationally invariant and isotropic (which, of course, is true for the fully disordered state), then Eq.\ \eqref{eq1} can be mapped onto the {\it continuum} diffusion equation, which in spherical coordinates reads
\begin{equation}
  \frac{\partial}{\partial t} R = \frac{\partial^2}{\partial r^2} R + (d-1) \frac{1}{r} \frac{\partial}{\partial r} R ,
  \label{rad}
\end{equation} 
i.e., $R = R(r,t)$, where $r$ is the radial coordinate and the boundary condition, $R(r = \varepsilon, t) = 1$, \mbox{$\varepsilon > 0$}, is imposed. Note that Eq.\ \eqref{rad} describes the behavior of the correlation function $C_{xy}$ {\it exactly} in the limit of large separations in arbitrary spatial dimensions $d$. Solving this equation is a standard exercise in statistical physics \cite{Krapivsky:2010}. At late times, one obtains
\begin{equation}
  R(\varepsilon < r < \xi(t)) \sim \left\{ \begin{matrix} 1 - r / \xi(t) ,& \quad d = 1 , \\[2pt]
    1 - \ln\left(r/\varepsilon\right) / \ln\left(\xi(t)/\varepsilon\right) ,& \quad d = 2 , \end{matrix} \right.
\end{equation}
where $\xi(t)\sim t^{1/2}$, while
\begin{equation}
  R(r > \varepsilon,t) \sim (\varepsilon/r)^{d-2} , \quad d > 2 .
\end{equation}
Thus, asymptotically the system develops long-range correlations. For $d\leq 2$ we find $R(r, t\rightarrow \infty) = 1$, while correlations decay algebraically for $d > 2$. We may estimate the timescale $t_N$ on which the system reaches the final, long-range correlated state by requiring $(1/N)\sum_{x,y}C_{xy}\sim \lim_{\varepsilon\rightarrow 0} \int_{\varepsilon}^{\xi(t)} dr \, r^{d-1} R(r,t) \sim N$, which yields
\begin{equation}
  t_N \sim \left\{ \begin{matrix} ~ N^2 ,& \quad d = 1 , \\
    ~ N \hspace{1pt} \ln N ,& \quad d = 2 , \\
    ~ N ,& \quad d > 2 . \end{matrix} \right.
    \label{scaling}
\end{equation}
These results are confirmed explicitly by solving numerically for the ensemble-averaged two-point function $C_{xy}$ on the hypercubic lattice starting from the fully disordered state \cite{Caspar:2015ddt}. Similarly, the spectral gap $\delta$ of the infinitesimal generator of the quantum dynamical semigroup $\boldsymbol{\mathcal{G}}$ scales as $\delta\sim t_N^{-1}$. The vanishing of $\delta$ in the limit $N\rightarrow\infty$ signals the presence of a diverging time- and lengthscale, $t_N\sim \xi^2$. This means that the system becomes critical in the infinite-volume limit.

{\it Competing dissipative couplings.} As a second example we consider a model that was introduced recently \cite{Weimer:2017} to investigate the possibility of purely dissipation-driven phase transitions in lattice quantum spin systems with Hamiltonian $H = 0$. The non-Hermitian jump operators $\ell$ are now taken from two distinct sets of bilocal operators, i.e., $\mathcal{S}_1 = \big\{ \big(L_1^{\alpha}\big)_{xy} \big\}$ with
\begin{equation}
  \big(L_1^{\alpha}\big)_{xy} = \frac{1}{2} \delta_{\langle x,y \rangle} \begin{pmatrix} \sqrt{2} s_{x}^{+} s_{y}^{+} (s_x^{-} - s_y^{-}) \\[5pt] 
  ~ -\sqrt{2} s_{x}^{-} s_{y}^{-} (s_x^{+} - s_y^{+}) ~ \\[5pt] 
  (s_{x}^{+}+s_{y}^{+} ) (s_x^{-} - s_y^{-} ) \end{pmatrix} ,
\end{equation}
and $\mathcal{S}_2 = \big\{ (L_2^{\beta})_{xy} \big\}$, with
\begin{equation}
  \big(L_2^{\beta}\big)_{xy} = \frac{1}{2} \delta_{\langle x,y \rangle} \begin{pmatrix} ( \mathbbm{1} + 2 s_x^{3} )  s_y^{-} \\[5pt]
    s_x^{-} ( \mathbbm{1} + 2 s_y^{3} ) \\[5pt]
    ~ s_x^{+} ( \mathbbm{1} - 2 s_y^{3} ) ~ \\[5pt]
    ( \mathbbm{1} - 2 s_x^{3} ) s_y^{+} \end{pmatrix} .
\end{equation}
The associated dissipative couplings are given by \mbox{$g_{\ell\in \mathcal{S}_1}\equiv g_1 = 1-\lambda$} and $g_{\ell\in \mathcal{S}_2}\equiv g_2 = \lambda$, with $\lambda\in [0,1]$, such that the infinitesimal generator can be decomposed into two distinct contributions $\boldsymbol{\mathcal{G}} = \boldsymbol{\mathcal{G}}_1 + \boldsymbol{\mathcal{G}}_2$, with $\boldsymbol{\mathcal{G}}_i\equiv g_i \sum_{\ell\in\mathcal{S}_i} \boldsymbol{\mathcal{D}}_{\ell}$. Note that the infinitesimal generator $\boldsymbol{\mathcal{G}}_1$ conserves $S^3$, i.e., $\boldsymbol{\mathcal{G}}_1 \big( S^3 \big) = 0$, while $\boldsymbol{\mathcal{G}}_2$ conserves $\widetilde{S}^3 = \sum_x (-1)^{||x||} s_x^3$. Nevertheless, the total generator for the dynamics $\boldsymbol{\mathcal{G}}$ conserves {\it neither} $S^3$ or $\widetilde{S}^3$, which makes it clear that the points $\lambda = 0$ and $\lambda = 1$ in parameter space are special. 

Again, we find that the hierarchy of correlation functions closes, arriving at an equation of the form \eqref{closed}. In particular, the evolution equations for the average magnetization $M = (1/N)\operatorname{tr} \rho_N S^3$ and the average staggered magnetization $M_{\rm s} = (1/N)\operatorname{tr} \rho_N \widetilde{S}^3$ are given by
\begin{align}
 \frac{d}{dt} M &=-8 d \lambda\hspace{1pt} M , \\
 \frac{d}{dt} M_{\rm s} &= -6 d (1-\lambda)\hspace{1pt} M_{\rm s} .
\end{align}
From the equations of motion, we confirm that the average (staggered) magnetization is conserved for $\lambda = 0$ ($\lambda = 1$). On the other hand, in the absence of conserved charges $0 < \lambda < 1$ both $M$ and $M_{\rm s}$ decay to zero, which characterizes a {\it unique} fixed point for the dynamics that is independent of the initial conditions.

\begin{figure}[t!]
  \centering
  \includegraphics[width=0.96\columnwidth]{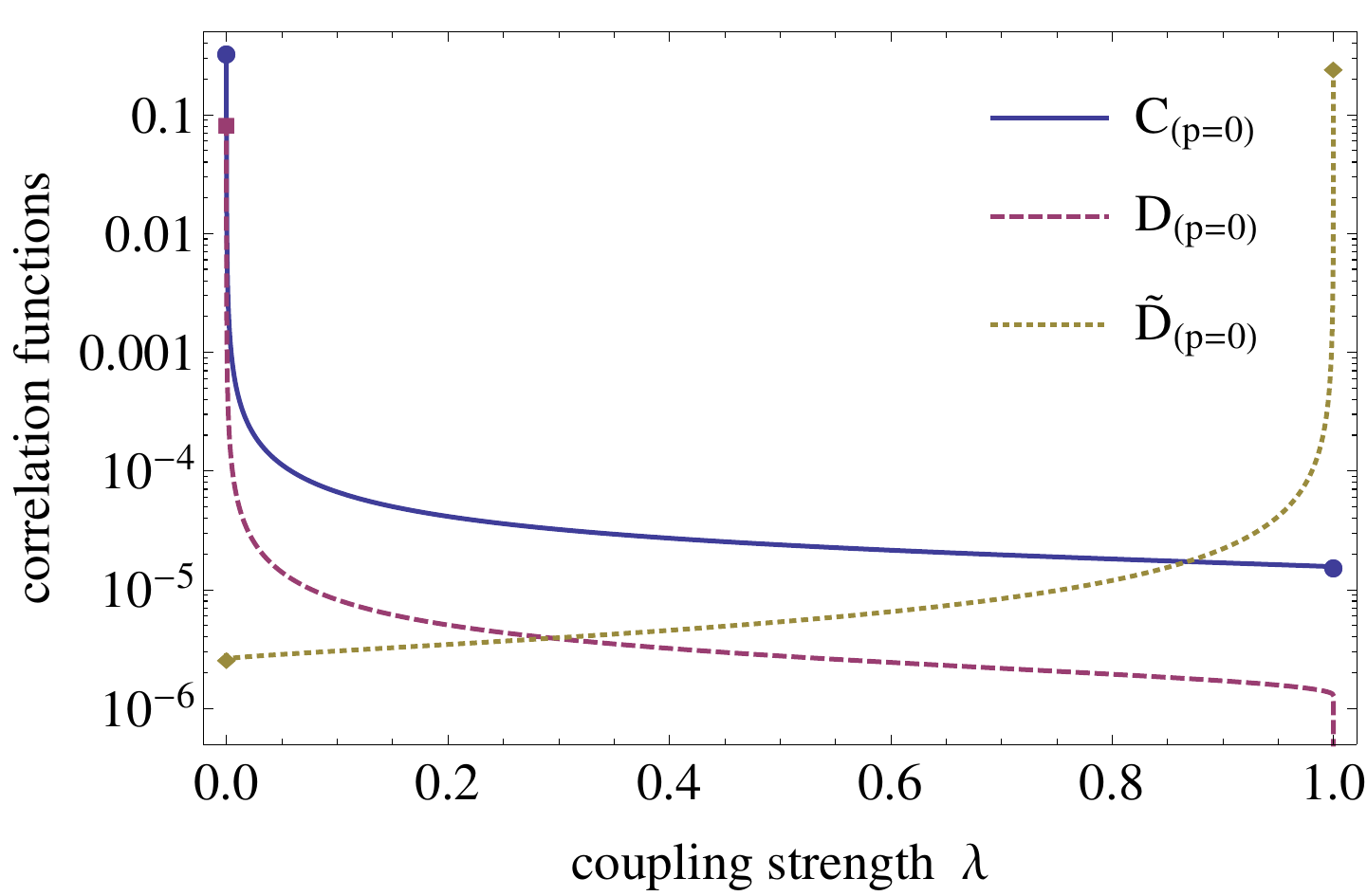}
  \caption{\label{fig1}Asymptotic values of the zero-modes of the correlation functions $C_{xy}$, $D_{xy}$ and $\widetilde{D}_{xy}$, e.g., $C_{(p=0)} = (1/N^2) \sum_{x,y} C_{xy}$, for $N = 40^3$ particles in $d = 3$ dimensions as a function of $\lambda$. Long-range correlations are observed infinitesimally close to the singular points $\lambda=0$ and $\lambda=1$.}
\end{figure}

\begin{figure}[!t]
  \centering
  \vspace{5pt}
  \includegraphics[width=0.9\columnwidth]{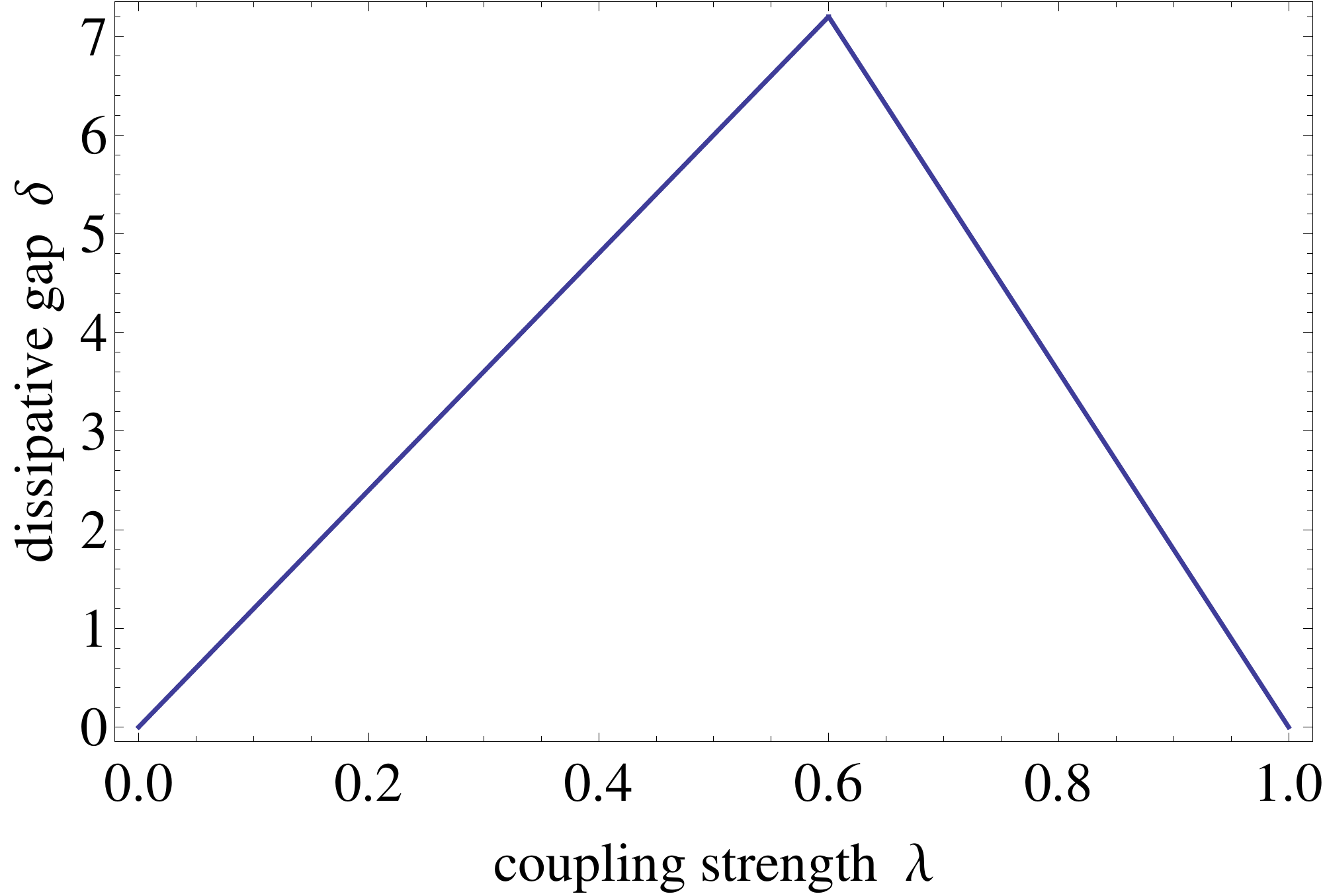}
  \caption{\label{fig2}Dissipative gap $\delta$ for a system of $N = 160^3$ particles in $d=3$ dimensions as a function of $\lambda$. The dissipative gap is finite for {\it all} $\lambda\in(0,1)$ while it scales to zero only in the limit $N\rightarrow \infty$ at the singular points $\lambda=0$ and $\lambda=1$.}
\end{figure}

The equations of motion for the ensemble-averaged two-point correlation functions $C_{xy}$ and $D_{xy}$ (defined above Eq.\ \eqref{eq1} and \eqref{eq2}), are given by ($x\neq y$)
\begin{align}
  \frac{d}{dt} C_{xy} &= \frac{3}{2} (1-\lambda) \Delta_{xy} C_{xy} - 8 d \lambda C_{xy} \notag \\
  & \hspace{10pt} -\delta_{\langle x,y\rangle} \left[ (1-\lambda)(C_{xx}+C_{xy}+8D_{xy}) - 4\lambda C_{xy} \right] ,  \\
  \frac{d}{dt} D_{xy} &= \frac{1}{2}(3-7\lambda) \Delta_{xy} D_{xy} - 16 d\lambda D_{xy}\notag \\
  & \hspace{10pt} - (1-\lambda)\delta_{\langle x,y\rangle} \left[ D_{xx} - D_{xy} + C_{xy} \right] ,
  \label{eqm2}
\end{align}
with $C_{xx} = 4 D_{xx} = 1$. It is useful to employ a different choice of basis and define the two-point function $\widetilde{D}_{xy} = (-1)^{||x - y||}\operatorname{tr}\rho_N s_x^3 s_y^3$. This function signals the presence of antiferromagnetic correlations and its equation of motion ($x\neq y$) reads
\begin{align}
\frac{d}{dt} \widetilde{D}_{xy} &= \frac{1}{2}(7\lambda - 3) \Delta_{xy}\hspace{1pt} \widetilde{D}_{xy} - 12 d (1-\lambda) \widetilde{D}_{xy}\notag \\
  & \hspace{10pt} + (1-\lambda)\delta_{\langle x,y\rangle} \left[ \widetilde{D}_{xx} + \widetilde{D}_{xy} + C_{xy} \right] ,
\end{align}
with $\widetilde{D}_{xx}\equiv D_{xx}  = 1/4$. Initializing the system in the fully disordered initial state $\rho_N=2^{-N}\mathbbm{1}$, we determine the presence of long-range correlations in the nonequilibrium steady state (Fig.\ \ref{fig1}), and calculate the characteristic timescales associated with the growth of correlations induced by the ``engineered'' dissipative reservoir (Fig.\ \ref{fig2}). This problem is discussed separately for two distinct cases, i.e., either $\lambda\in \{0, 1\}$ or $0 < \lambda < 1$. In the former case, the infinitesimal generator $\boldsymbol{\mathcal{G}} = \sum_{\ell} g_{\ell}\boldsymbol{\mathcal{D}}_{\ell}$ conserves either $M$ ($\lambda = 0$) or $M_{\rm s}$ ($\lambda = 1$) and the finite-size scaling of $t_N$ is identical to that derived in \eqref{scaling}. This is clear from the diffusive nature of the dynamics, which is ultimately a consequence of the presence of a (global) conserved charge. Long-range correlations in $C_{xy}$ ($\widetilde{D}_{xy}$) are hence only observed infinitesimally close to the special points $\lambda=0$ and $\lambda=1$ (cf.\ Fig.\ \ref{fig1}). On the other hand, for $0 < \lambda < 1$, the long-time behavior is fully determined by the presence of the {\it local} relaxation. Consequently, the characteristic time scale $t_N$ and the dissipative gap $\delta$ are {\it finite} for all $N$ (cf.\ Fig.\ \ref{fig2}). This explicitly rules out the possibility of a driven-dissipative phase transition between the two distinct long-range correlated phases at $\lambda = 0$ and $\lambda = 1$, which has been suggested previously \cite{Weimer:2017}. However, we want to point out that the dissipative gap $\delta$ nevertheless displays a nonanalytic behavior at $\lambda = 0.6$, which indicates a dynamic transition between a regime where the late-time dynamics is characterized either by the set of jump operators $\mathcal{S}_1$ or $\mathcal{S}_2$.

{\it Conclusions.} We have shown that closed hierarchies of correlation functions provide a unique perspective on the properties of open quantum spin systems. They allow us to reach definite conclusions regarding the vanishing of the dissipative gap, which has been suggested as a signature for steady-state phase transitions. In this way, specific proposals for quantum-state preparation via ``engineered'' dissipation can be solved exactly on large system sizes. Furthermore, this allows us to directly contrast our results to those of other numerical methods as, e.g., functional integral methods \cite{Hebenstreit:2016} or variational methods \cite{Haegeman:2011,Transchel:2014,Weimer:2015a,Weimer:2015b}, which can be tested on large system sizes in arbitrary dimensions. In particular, having discussed a recently introduced purely dissipative model for quantum magnetism \cite{Weimer:2017}, we find no indications of a nonequilibrium steady-state phase transition, which was suggested by variational calculations. We want to point out however that this does {\it not} exclude the possibility of singular (or nonanalytic) behavior in systems that exhibit closed hierarchies (see, e.g., \cite{Linzner:2016}).

In closing, we want to mention related work on ``solvable'' models in which {\it local} jump operators compete with the Hamiltonian dynamics \cite{Foss:2017}. While these models feature a closed hierarchy in the sense described in this work, they are different from the {\it purely dissipative} quantum spin models considered here, where jump operators act on {\it pairs} of neighboring lattice sites. Nevertheless, these two examples demonstrate that closed hierarchies provide a much broader framework for classifying and addressing ``solvable'' models with increasing computational complexity in a systematic way. It will certainly be interesting to study more elaborate closure conditions and their physical realizations in future work.

{\it Acknowledgments.} We want to thank S.\ Diehl, M.\ Fleisch\-hauer, \mbox{I.\ Lesanovsky}, and U.-J.\ Wiese for illuminating discussions. This research is funded by the Swiss National Science Foundation and the European Research Council under the European Union's Seventh Framework Programme, FP7/2007-2013, 339220.

\bibliography{references}

\end{document}